\documentclass[aps, prl, twocolumn, tightenlines, nobibnotes, notitlepage, superscriptaddress, nofootinbib, preprintnumbers, floatfix,showkeys,11pt,altaffilletter]{revtex4-2}

\usepackage{upgreek}

\usepackage[official]{eurosym}
\usepackage[normalem]{ulem}
\usepackage{amstext}
\usepackage{graphicx}
\graphicspath{{plots}}
\usepackage{url}
\usepackage{color}
\usepackage{ulem}
\usepackage[version=4]{mhchem}
\usepackage[utf8]{inputenc}
\usepackage{fontawesome}
\usepackage{yfonts}
\usepackage{slashed}

\pdfoutput=1
\usepackage{textcomp}
\usepackage{comment}
\usepackage{yfonts}
\usepackage{epsfig,amsfonts,mathrsfs,graphicx,color,slashed,multirow}
\usepackage{latexsym,graphicx,slashed,color,enumerate,url,cancel,gensymb}
\usepackage{textcomp}

\usepackage[x11names]{xcolor}
\usepackage[colorlinks,pdfstartview=FitV,breaklinks=true]{hyperref}
\usepackage{booktabs}
\usepackage{adjustbox}
\usepackage{textgreek} 
\usepackage{booktabs} 
\usepackage{float}

\usepackage{lmodern}
\usepackage{ae,aecompl}
\usepackage{appendix}
\usepackage{orcidlink}
\usepackage{nccmath} 
\usepackage{textcomp}
\allowdisplaybreaks 

\newcommand{\cevns}{CE$\nu$NS}
\newcommand{\Cr}{$^{51}$Cr}
\newcommand{\Ar}{$^{37}$Ar}
\newcommand{\G}{$\gamma$}
\newcommand{\A}{$\alpha$}
\newcommand{\lwo}{Li$_2$WO$_4$}


\def\cevns{CE\textnu NS}

\newcommand{\qtransfer}{\left|\mathbf{q}\right|}

\definecolor{byzantium}{rgb}{0.44, 0.16, 0.39}

\makeatletter
    \newcommand{\colorboxed}[3][white]{\fcolorbox{#2}{#1}{\m@th$\displaystyle#3$}}
\makeatother

\AtBeginDocument{\hypersetup{citecolor=byzantium,linkcolor=byzantium,urlcolor=byzantium}}
\usepackage{appendix}

\begin{document}

\title{\LARGE Prospects for precision \cevns\ measurements with electron-capture neutrinos and lithium-based bolometers}

\author{Giovanni Benato~\orcidlink{0000-0002-8768-290X}}
\email{giovanni.benato@gssi.it}
\affiliation{INFN – Laboratori Nazionali del Gran Sasso, Assergi (L’Aquila) I-67100, Italy
}
\affiliation{Gran Sasso Science Institute, L’Aquila I-67100, Italy}

\author{Francesca M. Pofi~\orcidlink{0009-0003-8906-6072}}
\email{francescamaria.pofi@gssi.it}
\affiliation{INFN – Laboratori Nazionali del Gran Sasso, Assergi (L’Aquila) I-67100, Italy
}
\affiliation{Gran Sasso Science Institute, L’Aquila I-67100, Italy}

\author{Andrei Puiu~\orcidlink{0000-0003-1143-224X}}
\email{andrei.puiu@lngs.infn.it}
\affiliation{INFN – Laboratori Nazionali del Gran Sasso, Assergi (L’Aquila) I-67100, Italy
}
\author{Christoph A. Ternes~\orcidlink{0000-0002-7190-1581}}
\email{christoph.ternes@lngs.infn.it}
\affiliation{INFN – Laboratori Nazionali del Gran Sasso, Assergi (L’Aquila) I-67100, Italy
}
\affiliation{Gran Sasso Science Institute, L’Aquila I-67100, Italy}


\begin{abstract}
  We evaluate the feasibility of high-precision coherent elastic neutrino-nucleus scattering measurements
  exploiting mono-energetic neutrinos produced by electron-capture (EC) decays of intense radioactive sources, such as \Cr\ or \Ar.
  To fully exploit the high neutrino flux achievable with EC sources, and accounting for the low energy of EC neutrinos,
  we evaluate the sensitivity of a compact array of bolometric detectors with absorbers based on light elements, such as lithium, oxygen and fluorine.
  With 1\,kg of LiF detectors, source activities of $\sim10^{17}$\,Bq, and assuming an energy threshold of 20\,eV,
  we expect to achieve a $\sim$3\% precision on the determination of the neutrino flux with 90\, days of measurement.
  Such a measurement would represent a test of the gallium neutrino anomaly,
  and could disentangle nuclear effects on gallium from the misevaluation of the EC source activity
  or short-baseline neutrino oscillations.
\end{abstract}

\maketitle

Coherent elastic neutrino nucleus scattering (\cevns) is a Standard Model process predicted in 1973~\cite{Freedman:1973yd}
and first observed in 2017 on a CsI target~\cite{COHERENT:2017ipa} by the COHERENT collaboration.
More recently, COHERENT has strengthened the CsI result~\cite{COHERENT:2021xmm}
and measured \cevns\ also on argon~\cite{COHERENT:2020iec} and germanium targets~\cite{COHERENT:2024axu}.
\cevns\ has also been observed using solar neutrinos on xenon targets in XENONnT~\cite{XENON:2024ijk}, PandaX-4T~\cite{PandaX:2024muv} and LUX-ZEPLIN~\cite{LZ:2025igz},
and reactor antineutrinos on germanium targets in CONUS+~\cite{Ackermann:2025obx}. 
Previously, a measurement was performed using antineutrinos from the DRESDEN-II power plant~\cite{Colaresi:2022obx},
even though this result is disputed because of its incompatibility with other measurements~\cite{CONUSCollaboration:2024kvo,TEXONO:2024vfk,nGeN:2025hsd,Ackermann:2025obx,Li:2025pfw}.

The differential cross section of the \cevns\ process, with respect to the nuclear recoil energy $T_\mathcal{N}$ is given by~\cite{Freedman:1973yd,Barranco:2005yy}
\begin{align}
  \frac{d\sigma_{\nu \mathcal{N}}}{dT_\mathcal{N}}=&\frac{G_F^2 m_\mathcal{N}}{\pi}\left({Q_V^\mathrm{SM}}\right)^2 F_{W}^2(\qtransfer^2)\nonumber\\
  \times&\left(1-\frac{m_\mathcal{N} T_\mathcal{N}}{2E_\nu^2}\right)\,,
  \label{eq:xsec_CEvNS_SM}
\end{align}
where $G_F$ is the Fermi constant, $E_\nu$ the neutrino energy, $m_\mathcal{N}$ the nuclear mass, and $Q_V^\text{SM}$ the weak charge given by
\begin{equation}
\label{eq:CEvNS_SM_Qw}
    Q_V^\text{SM} = g_V^p Z + g_V^n N\,,
\end{equation}
where $Z$ and $N$ are the proton and neutron numbers, respectively. 
The proton and neutron couplings at tree level are given by $g_V^p = (1- 4 \sin^2 \theta_W)/2$ and $ g_V^n = -1/2$,
where $\sin^2 \theta_W=0.23857(5)$ is the value of the weak mixing angle at low energies~\cite{ParticleDataGroup:2024cfk}. 
At higher orders these factors become flavor-dependent. While the correction to $g_V^n$ is very small,
the correction to $g_V^p$ is quite significant; see the discussions in Refs.~\cite{AtzoriCorona:2024rtv,AtzoriCorona:2025xwr}.
Corrections due to nuclear physics are included in the form factor $F_{W}^2(\qtransfer^2)$, to account for
the finite nuclear spatial distribution, for which we will use in our analyses the Klein-Nystrand parameterization~\cite{Klein:1999qj}. 
Note, however, that due to the small energies and small nuclear masses relevant here, $F_{W}^2(\qtransfer^2)\simeq 1$.

In the present work, we discuss the requirements and evaluate the sensitivity of a possible measurement
of \cevns\ using neutrinos produced by a \Cr\ or \Ar\ electron-capture (EC) source.
\Cr\ decays to the $^{51}$V ground state emitting 747\,keV neutrinos with a branching ratio of 90.07\%,
and to a $^{51}$V excited state emitting 427\,keV neutrinos (followed by the emission of a 320\,keV \G) with 9.93\% branching ratio.
Instead, \Ar\ always decays to the $^{37}$Cl ground state emitting 811\,keV neutrinos.
The half-life values are 27.7 and 35.0 days for \Cr\ and \Ar, respectively.
\Ar\ is favorable because of the slightly higher neutrino energy and the absence of emitted \G's,
which significantly relaxes the required thickness of the source shielding.
On the other hand, the source production and dimensions need also to be considered.
So far, \Ar\ has been produced via $^{40}$Ca(n,\A)\Ar\ in fast-neutron reactors~\cite{Barsanov:2007fe}.
This method has demonstrated the possibility to reach an activity of $1.5\times10^{16}$\,Bq,
but requires the purification of the irradiated target.
An alternative production channel is the thermal neutron capture on a $^{36}$Ar-enriched argon target~\cite{Baudis:2020nwe,Zhang:2024ucs},
which would avoid any post-processing of the irradiated material at the price of handling a high-pressure gas bottle next to a reactor core.
Similarly, \Cr\ can be produced via thermal neutron capture on $^{50}$Cr-enriched chromium, as demonstrated in GALLEX~\cite{Cribier:1996cq},
SAGE~\cite{SAGE:1998fvr} and BEST~\cite{Danshin:2022wiz}.
The cross-sections are $\sim5$\,b and $\sim15$\,b for $^{36}$Ar and $^{50}$Cr, respectively~\cite{Pritychenko:2025bqr,osti_2571019_endfbviii1}.
The highest currently available thermal neutron flux is achieved by the ILL High-Flux Reactor,
which reaches $1.5\times10^{15}$\,n/cm$^2$/s, operating in cycles of 50-60 days~\cite{Bergeron2010RHFLEU}.
Assuming the irradiation of 1\,kg of isotope for a full reactor cycle, the attainable source activity
is $8.6\times10^{16}$\,Bq for \Ar, and $2.0\times10^{17}$\,Bq for \Cr.
We will consider these values for the sensitivity evaluation, even though
higher activities are attainable via the irradiation of larger target masses.

\begin{figure}[htbp]
\centering
\includegraphics[width=0.49\textwidth]{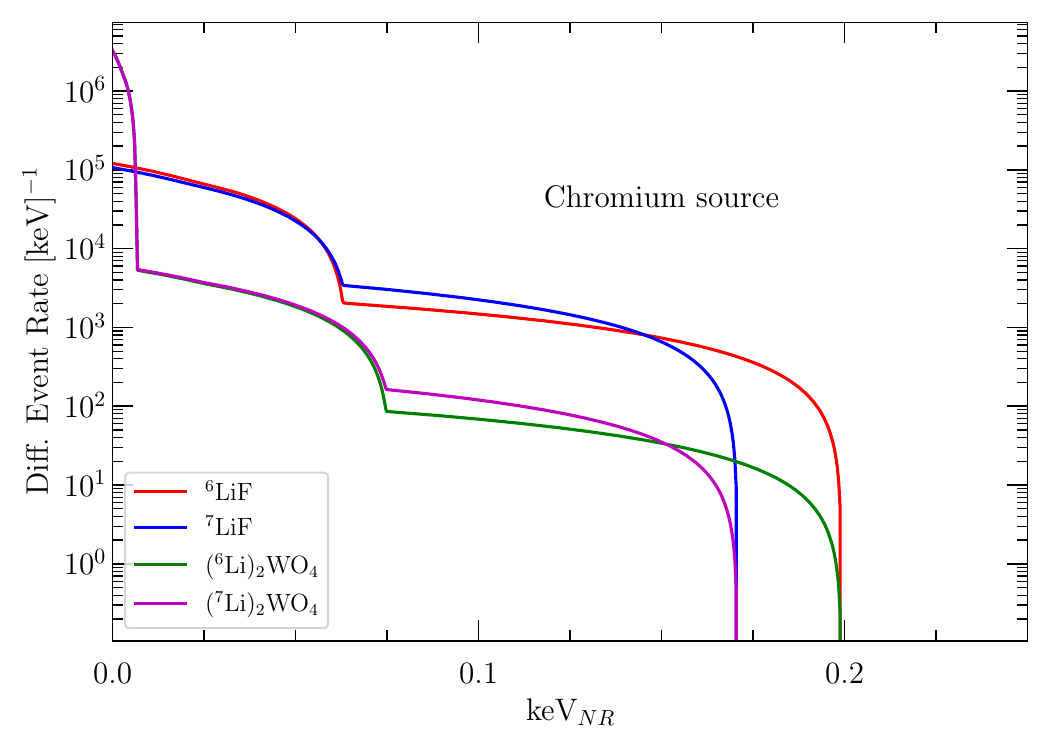}
\\
\includegraphics[width=0.49\textwidth]{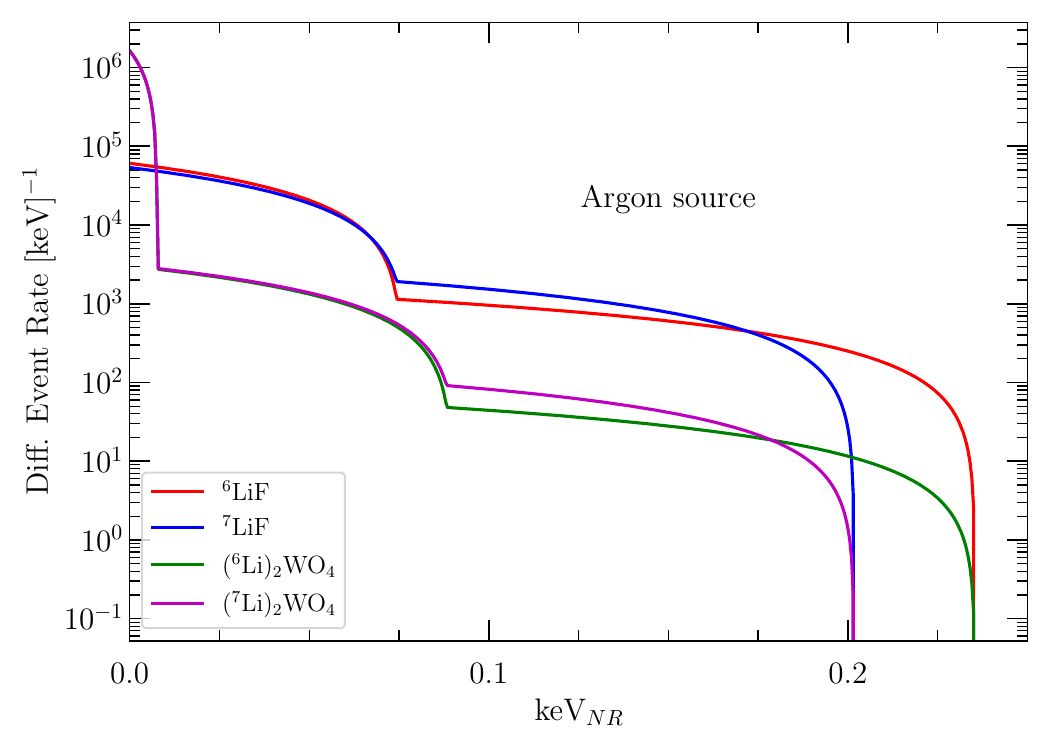}
\caption{Expected event rate for different lithium-based crystals for a 1\,kg detector and 90 days exposure using a \Cr\ (upper panel) and \Ar\ (lower panel) source.}
\label{fig:eventrate}
\end{figure}

The low energy of EC neutrinos imposes the use of detectors with a trigger threshold at the eV scale~\cite{Bellenghi:2019vtc},
practically restricting the choice to cryogenic calorimeters with absorbers made of light atoms to achieve larger nuclear recoil signals,
at the cost of a reduced event rate.
In particular, we are considering the detection of \cevns\ on lithium, oxygen and fluorine using LiF or \lwo\ crystals~\cite{Helis:2025cur},
which allows to relax the threshold requirement to $\sim$20\,eV.
An advantage of lithium is that it can be easily enriched in $^{6}$Li or $^{7}$Li,
allowing us to split the detectors in two types and exploit the different detected count rate
to cross-check the detected neutrino activity.
We envisage a setup with a cylindrical array of detectors surrounding the source at a minimum distance of $\sim$12~cm.
For \Cr, this distance allows for the installation of the source embedded in a $\sim$8\,cm thick tungsten shield,
while in the case of \Ar, it leaves enough space for the insertion of a spherical argon bottle at 300\,bar of pressure.
We assume a geometrical efficiency of 50\% throughout our calculations.
Larger detector masses would increase the geometrical efficiency, hence this choice is rather conservative.
Assuming a 1~kg detector and an exposure of 90 days to account for the decay of the source,
we obtain the expected event rates shown in Fig.~\ref{fig:eventrate}.
We assume the lithium to be 100\% enriched to either $^6$Li or $^7$Li, as indicated in the legend.
Several kinks can be observed in the figure,
corresponding to the various isotopes contained in LiF and \lwo. In particular, the components with the highest end-points correspond to $^6$Li and $^7$Li.
In the case of \lwo, \cevns\ on tungsten would become relevant for extremely low energies $T_\mathcal{N}\lesssim 7$~eV,
which are below the energy threshold considered here. 
We highlight that the reported mass refers to the detector material:
since 1\,kg of LiF crystals contains a higher number of Li targets than the same amount of \lwo,
the event rates for the different crystals differ even above $\sim80$\,eV, where only \cevns\ on lithium is relevant.

In the following, we evaluate the detector size required to achieve a measurement of the neutrino flux from the \Ar\ and \Cr\ sources
at a given precision.
In all cases, we will assume an exposure of 90 days, corresponding to roughly three times the half-life of \Cr\ or \Ar.
We assume that the activity of the source is known with a percent precision~\cite{Barsanov:2007fe,Cribier:1996cq,SAGE:1998fvr}, making the uncertainty on the activity negligible.
In addition, we assume to operate in a background-free regime:
any real experiment would be affected by a backgrounds, but any estimate would inevitably be arbitrary at this stage.
In particular, the Low-Energy-Excess (LEE)~\cite{Fuss:2022fxe,Baxter:2025odk} would be a major technical obstacle
that would have to be solved to allow for the actual implementation of the idea presented here.

\begin{figure}[htbp]
\centering
\includegraphics[width=0.49\textwidth]{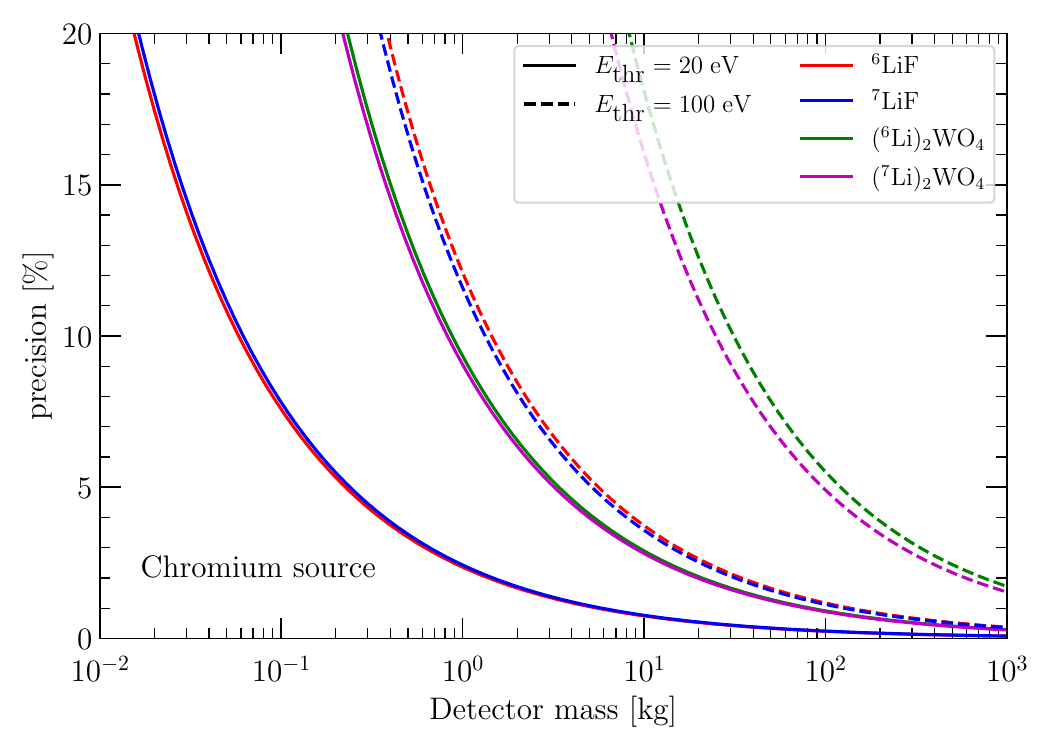}
\\
\includegraphics[width=0.49\textwidth]{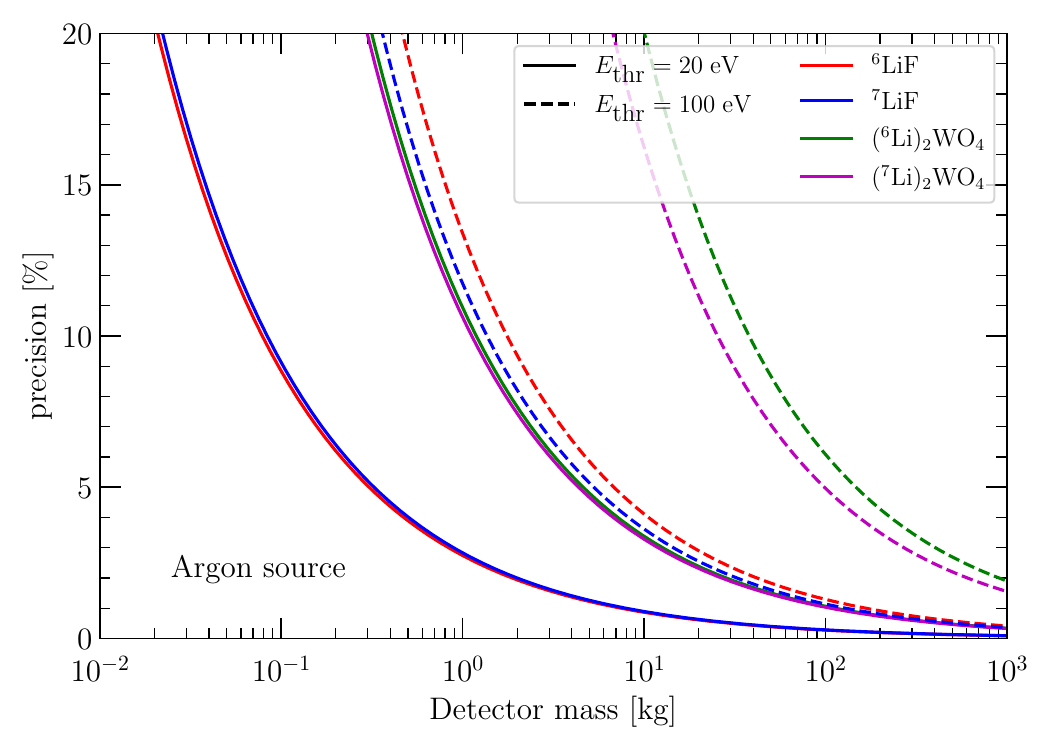}
\caption{Attainable precision on the neutrino flux as a function of the detector mass for different crystals,
  assuming 90~days of exposure and with energy thresholds of 20\,eV (solid lines) or 100\,eV (dashed lines).
  The upper (lower) panel shows the precision achievable with the \Cr\ (\Ar) source.}
\label{fig:mass_estimate}
\end{figure}

Figure~\ref{fig:mass_estimate} shows that, assuming an energy threshold of 20~eV,
a $\sim1$\,kg LiF detector or a $\sim10$\,kg \lwo\ detector could already yield a $\sim$3\% precision on the neutrino flux,
allowing to directly test the Gallium neutrino anomaly on a different detection channel than the one investigated so far, as discussed below.
A similar precision could be achieved with $\sim10$\,kg of LiF in the more conservative scenario of a 100\,eV threshold.
The attainable precision in the presence of the LEE would need to be evaluated as a function of the LEE amplitude.
However, we point out that the experimental approach presented here features a few significant improvements with respect to the strategies followed so far.
First, EC sources yield high fluxes of mono-energetic neutrinos, and simultaneously bypass the two main limiting factors
that could prevent the use of \cevns\ for precision measurements or Beyond the Standard Model searches,
i.e. the low flux of accelerator neutrinos, and the limited precision affecting the spectral shape of reactor neutrinos.
Second, the combined use of $^6$Li- and $^7$Li-enriched detectors would allow a cross-validation
of the measured neutrino flux. In addition, the use of such light elements could open the possibility of a measurement of the axial-vector coupling constant with \cevns, see also Ref.~\cite{AristizabalSierra:2026rlo}.
Finally, considering the technological improvement demonstrated by the cryogenic calorimeter community over the recent past,
we can expect that in the coming decades cryogenic detectors with masses of the order of tens of grams
and energy threshold at the eV level could be conceivable, allowing the use of targets composed of mid- or high-mass elements,
which would yield a much higher event rate per unit mass if compared to lithium, oxygen or fluorine.

A first goal for the experimental approach presented here could be the test of the Gallium neutrino anomaly,
which was originally a deficit of events observed in the GALLEX~\cite{GALLEX:1994rym,GALLEX:1997lja,Kaether:2010ag}
and SAGE~\cite{Abdurashitov:1996dp,SAGE:1998fvr,Abdurashitov:2005tb,SAGE:2009eeu} source experiments. 
Two source measurements have been performed by the GALLEX collaboration using an intense artificial \Cr\ radioactive source placed inside the detector.
Instead, the SAGE collaboration performed one measurement with a \Cr\  source and another one with a \Ar\ source.
The significance of the anomaly was originally at the 2--3$\sigma$ level,
depending on the exact cross-section model considered in the calculation~\cite{Abdurashitov:2005tb,Laveder:2007zz,Giunti:2006bj,Kostensalo:2019vmv}. 
Later on, the anomaly was confirmed in the BEST source experiment~\cite{Barinov:2021asz,Barinov:2022wfh},
pushing the significance to the 5--6$\sigma$ level~\cite{Giunti:2022btk,Berryman:2021yan,Barinov:2021mjj}. 
The Gallium anomaly could be, in principle, explained with short-baseline oscillations due to light sterile neutrinos
(see the reviews in Refs.~\cite{Gariazzo:2015rra,Gonzalez-Garcia:2015qrr,Giunti:2019aiy,Diaz:2019fwt,Boser:2019rta,Dasgupta:2021ies}). 
However, the parameter space required to explain the 20\% deficit obtained from the combined analysis of GALLEX, SAGE and BEST data
is in tension with measurements performed by other experiments~\cite{Giunti:2022btk,Giunti:2021kab,Berryman:2019hme,Gonzalez-Garcia:2024hmf}. 
Several other oscillation explanations~\cite{Giunti:2010zs,Forero:2022skg,Arguelles:2022bvt,Hardin:2022muu,Banks:2023qgd}
suffer from similar tensions with some other data~\cite{Tortola:2020ncu,Giunti:2022btk,Giunti:2023kyo,deGouvea:2024syg}. 
So far, non-standard decoherence effects are the only oscillation explanation that does not appear to have tensions~\cite{Farzan:2023fqa}. 
Explanations with Standard Model physics have been suggested in Refs.~\cite{Giunti:2022xat,Brdar:2023cms,Cadeddu:2025ueh}.
In particular it was suggested~\cite{Giunti:2022xat} that the Gallium anomaly could be resolved with a longer half life of $^{71}$Ge.
This possibility has already been excluded with new dedicated measurements, see Refs.~\cite{Collar:2023yew,Norman:2024hki}.
In Ref.~\cite{Cadeddu:2025ueh} it has been suggested that that the gallium anomaly could be explained
by recalculating the neutrino-capture cross section using a non-factorized treatment of the leptonic and nuclear wave functions.
In light of this yet unresolved situation, the idea proposed here could represent an independent cross-check of the gallium experiments,
and could allow testing whether the anomaly is indeed caused by a miscalculation
of the neutrino capture cross section on $^{71}$Ga, or if it is related to either the source activity or the neutrino propagation.
Other experiments to test the gallium anomaly have been proposed recently~\cite{Huber:2022osv,Chauhan:2025vly,Semenov:2026xfx}.
The advantages of the approach presented here lie in the compactness of the experimental design,
in the relatively limited cost of the experimental infrastructure, and in the much broader physics reach
that such an experiment could provide.

\begin{figure}[htbp]
\centering
\includegraphics[width=0.49\textwidth]{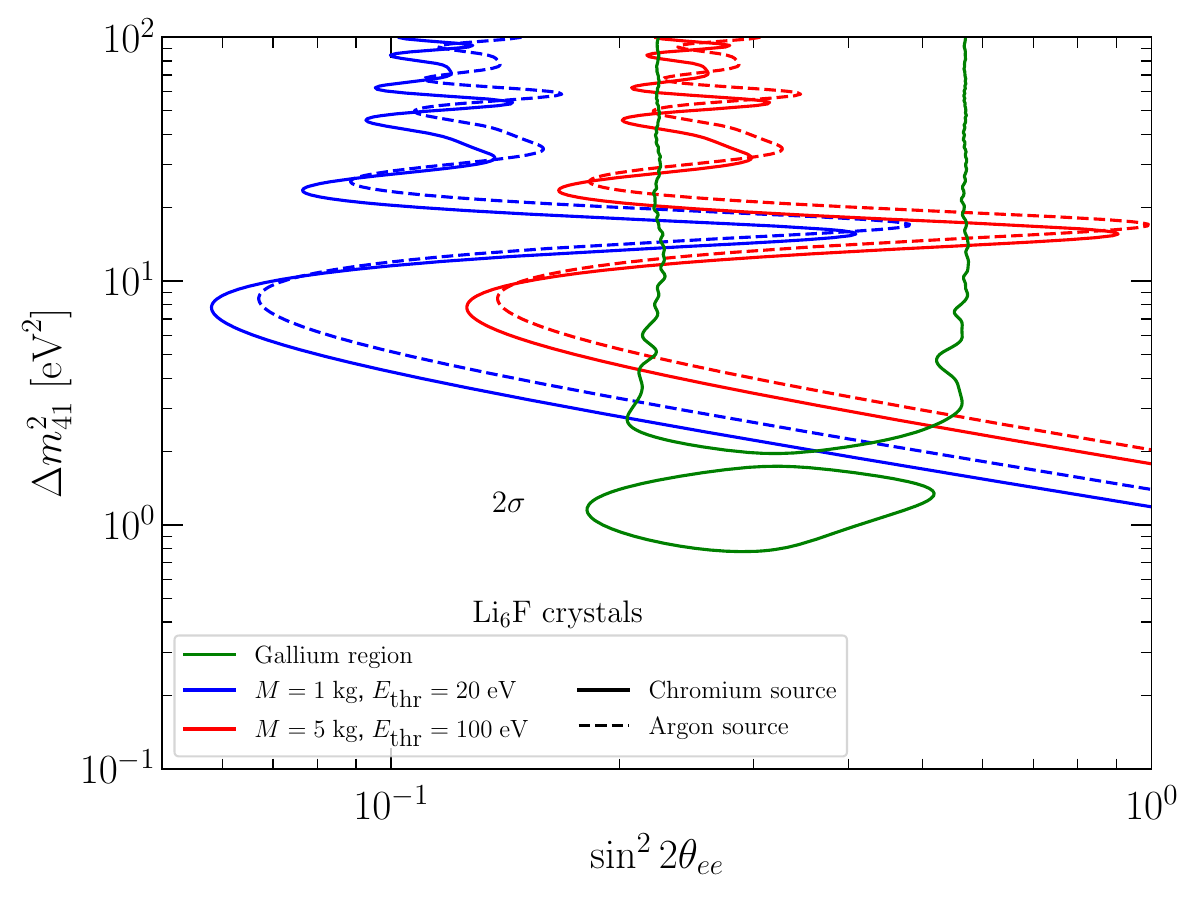}
\caption{Exclusion sensitivity contours at 2$\sigma$ for light sterile neutrino oscillations
  assuming 90~days of measurement using a 1\,kg detector with 20\,eV threshold (blue)
  or a 5\,kg detector with 200\,eV threshold (red).
  The solid and dashed lines correspond to the use of a \Cr\ or \Ar\ source, respectively.
  The green line corresponds to the parameter space allowed to explain the gallium neutrino anomaly
  with neutrino oscillations.}
\label{fig:gallium}
\end{figure}

In Fig.~\ref{fig:gallium} we show the sensitivity to test the parameters characterizing neutrino oscillations due to light sterile neutrinos,
\begin{equation}
    P_{ee} = 1-\sin^22\theta_{ee}\sin^2\left(\frac{\Delta m_{41}^2L}{4E}\right)\quad.
\end{equation}
Here we only consider $^6$LiF as detection material. The solid and dashed lines are obtained considering the \Cr\ and \Ar\ sources, respectively. 
We show in blue the expected 2$\sigma$ exclusion sensitivity for a 90 day exposure using a 1\,kg detector.
In green we show the parameter space required to explain the Gallium anomaly with neutrino oscillations due to light sterile neutrinos obtained with the Bahcall cross section model~\cite{Bahcall:1997eg}. 
We use this for illustration. Other cross section models~\cite{Haxton:1998uc,Kostensalo:2019vmv,Frekers:2015wga,Semenov:2020xea,Haxton:2025hye} lead to similar contours~\cite{Giunti:2022btk,Giunti:2022xat}. 
As can be seen with a single activation of the source and a 90 day measurement using a 1~kg detector most of the parameter space required to explain the Gallium anomaly would be tested.
Even with a larger energy threshold and using a 5\,kg detector an important part of the parameter space would be excluded as indicated by the red lines.
The wiggles in the exclusion curves are connected to the relavive distance between the source and the detector,
and could be removed by repeating the measurement with the source at different distances.

We would like to highlight that the experimental concept proposed in this work spans well beyond the search for light sterile neutrinos,
and could open the field to precision measurement of \cevns\ using light nuclei.
The test of the Gallium neutrino anomaly is simply a straightforward application that could be achieved with a small-scale implementation of the detector concept.
It is an important result, since it would present an independent test of our understanding of the sources used in previous experiments. 
Our experimental approach could prove useful for flux determinations, precision measurements, and searches for physics beyond the Standard Model,
which often manifest at low recoil energies accessible at our proposed experiment.


\bibliographystyle{utphys}
\bibliography{bibliography}  

\end{document}